\documentclass[aps,pra,floats,floatfix,twocolumn]{revtex4}

\usepackage{ifthen}
\usepackage{ifpdf}
\usepackage{color}

\ifpdf
\usepackage{graphicx}
\usepackage{epstopdf}
\else
\usepackage{graphicx}
\usepackage{epsfig}
\fi

\usepackage{latexsym}
\usepackage{amsmath}
\usepackage{amssymb}
\usepackage{bm}
\usepackage{wasysym}

\newcommand{\eexp}{\mbox{e}^}

\newcommand{\mylabel}[1]{\label{#1}} 
\newcommand{\beq}{\begin{eqnarray}}
\newcommand{\eeq}{\end{eqnarray}} 
\newcommand{\be}[1]{\begin{eqnarray}\ifthenelse{#1=-1}
{\nonumber}{\ifthenelse{#1=0}{}{\mylabel{e#1}}}}
\newcommand{\ee}{\end{eqnarray}} 

\newcommand{\hide}[1]{[hidden text]}
\renewcommand{\cite}[1]{\textcolor{blue}{[\onlinecite{#1}}]} 

\newcommand{\ei}{\hat{a}}
\newcommand{\eidag}{\hat{a}^{\dag}}
\newcommand{\hn}{\hat{n}}

\newcommand{\Jx}{\hat{J}_x}
\newcommand{\Jy}{\hat{J}_y}
\newcommand{\Jz}{\hat{J}_z}
\newcommand{\Ji}{\hat{J}_i}
\newcommand{\Jj}{\hat{J}_j}

\begin{document}

\title{Squeezing in driven bimodal Bose-Einstein Condensates: Erratic driving versus noise}

\author{Christine Khripkov$^1$, Amichay Vardi$^1$, and Doron Cohen$^2$}
\affiliation{Departments of $^1$Chemistry and $^2$Physics, Ben-Gurion University of the Negev, Beer-Sheva 84105, Israel}


\begin{abstract}
We study the interplay of squeezing and phase randomization 
near the hyperbolic instability of a two-site Bose-Hubbard model 
in the Josephson interaction regime.  
We obtain results for the quantum Zeno suppression 
of squeezing far beyond the previously found short time
behavior.  
More importantly, we contrast the expected outcome   
with the case where randomization is induced by erratic driving 
with the same fluctuations as the quantum noise source, finding significant differences. 
These are related to the distribution of the squeezing factor, 
which has log-normal characteristics: hence its average is significantly 
different from its median due to the occurrence of rare events. 
\end{abstract}  


\maketitle

\section{Introduction}
The effect of stochastic driving on unitary evolution has been a central theme of modern quantum mechanics. It is well established that quantum decay can be suppressed by frequent interventions, or measurements,  or by the introduction of noise, via the Quantum Zeno Effect (QZE) \cite{QZE1,QZE2,QZE3,QZE4,Kofman1,Kofman2,Kofman3,Gordon1,Gordon2,
AZE1,AZE2,AZE3}.  The modelling of the ``interventions" as arising from a {\em deterministic} or from a {\em noisy} source are often used interchangeably \cite{Kofman1,Kofman2,Kofman3,Gordon1,Gordon2}. This partially reflects the paradigm that the Langevin picture and the Master equation picture of the dynamics are equivalent.    

Recent work considered the QZE suppression of interaction-induced squeezing 
in bimodal Bose-Einstein condensates \cite{QZESQ1,QZESQ2}. Since matter-wave squeezing 
is the key to the realization of atom interferometers below the standard quantum 
limit \cite{SubShotNoise1,SubShotNoise2,SubShotNoise3,SubShotNoise4,
SubShotNoise5,SubShotNoise6}, it is highly desirable to gain better understanding 
of its interplay with noise.  Noise was shown to arrest the squeezing 
and build-up of many-body correlations in the large, multi-particle system, 
prepared with all particles occupying the odd superposition of the two-modes. 
In the Josephson regime \cite{Gati07} this preparation constitutes a hyperbolic saddle point, 
leading to a rapid squeezing \cite{Hyperbolic1,Hyperbolic2}. It was shown that the degree 
of squeezing and the associated phase diffusion \cite{PhaseDiffusion1,PhaseDiffusion2,PhaseDiffusion3,PhaseDiffusion4,
PhaseDiffusion5,PhaseDiffusion6} 
could be controlled by a noisy modulation of the coupling between the modes,  
up to a full arrest via a Bose-stimulated QZE \cite{QZESQ1,QZESQ2}.

%
In this work we attain two principle goals.
{\bf (i)}  
We extend the analytic understanding of the QZE suppression of squeezing 
to time-scales which are orders of magnitude longer than these of Ref.~\cite{QZESQ1,QZESQ2},  
obtaining good agreement with numerical simulations. 
{\bf (ii)}
We challenge the fundamental paradigm of replacing quantum noise by deterministic 
erratic driving. Erratic driving can have non-trivial {\em statistics}, 
hence its {\em typical} results do not have to agree with the {\em average} behavior.  
This is demonstrated in our system by an important caveat resulting from 
the interplay of the nonlinear squeezing dynamics and 
the diffusive randomization by driving. While the early evolution 
under the influence of either noisy or erratic driving 
corresponds to the QZE of Ref.~\cite{QZESQ1,QZESQ2}, significant differences 
arise at later times. These differences are explained by a statistical analysis: 
As the squeezing is hyperbolic, while the driving induces diffusion,  
the resulting stretch distribution has log-normal characteristics, 
with rare events separating its mean from its median. 
The outcome of a typical erratic driving scenario is likely to reflect the median, 
and might be significantly different from the outcome of a full Feynman-Vernon 
averaging that is required for the description of quantum noise. 
Using semiclassical reasoning \cite{Chuchem10} we derive 
analytic expressions for the median and for the mean single-particle coherence, 
given the known normal statistics of the squeezing parameter.

In Section II we present the model driven Bose-Hubbard system, the pertinent initial conditions, and the relation between the squeezing parameter and the observed fringe visibility for Gaussian squeezed states. The coherence dynamics with and without noise are presented in Section III. The concept of erratic driving and its statistical analysis are introduced in Section IV and a short summary is provided in Section V.

%
\section{Modelling}

We consider the dynamics generated by the two-mode Bose-Hubbard Hamiltonian
 (BHH) \cite{Gati07,Hyperbolic1,Hyperbolic2,Chuchem10} with an additional driving source,
\begin{equation}
\label{Ham}
\hat{H} \ = \ U\Jz^2  - [K+f(t)]\Jx~,
\end{equation}
where $\Jx=(\eidag_1 \ei_2+\eidag_2\ei_1)/2$, $\Jy=(\eidag_1\ei_2-\eidag_2\ei_1)/(2i)$, and $\Jz=(\hn_1 - \hn_2)/2$. The $\ei_i$ and $\eidag_i$ are bosonic annihilation and creation operators, respectively. 
The particle number operator in mode~$i$ is $\hn_{i}=\eidag_{i}\ei_{i}$. 
The total particle number $\hn_1+\hn_2=N=2j$ is conserved. 
The dimensionless interaction parameter is ${u=NU/K}$. We note that the undriven two-mode BHH is known in nuclear physics as the Lipkin-Meshkov-Glick model \cite{LMG,Ribiero}, and it has been also used to describe interacting spin systems \cite{Botet83} and magnetic molecules \cite{Garanin98}. Our interest lies in the Josephson regime where ${1\ll u \ll N^2}$ \cite{Gati07}, 
hence in the classical limit ${{\bf J}_0=(-j,0,0)}$ is a hyperbolic point. 
The driving source induces a fluctuating field $f(t)$
which corresponds to the modulation of the barrier in a double-well realization of the two-mode BHH.
We assume that this perturbation has a zero average and a short correlation time, 
such that upon averaging over time,  
\beq
\langle f(t) f(t') \rangle \ \ = \ \ 2D\delta(t-t')~.
\eeq
Hence the averaged dynamics is described by a Master equation
which includes a term that generates angular diffusion around 
the $J_x$ axis:
\begin{equation}
\frac{d}{dt}\hat{\rho} \ = \ -i\left[\hat{H},\hat{\rho}\right]-D\left[\Jx,\left[\Jx,\hat{\rho}\right]\right]~,
\label{master}
\end{equation}
where $\hat{\rho} $ is the $N$-particle density matrix. We preform numerical simulations of two possible scenarios: 
{\bf (a)} Dynamics that is generated by the master equation; 
{\bf (b)} Dynamics that is generated by a typical realization of $f(t)$ \cite{rA}.
Formally, the mixed state obtained in (a) can be 
regarded as the average over the pure states  
obtained in (b), provided that {\em all} possible realizations of $f(t)$ are included.      

\begin{figure}
\centering
\includegraphics[width=0.48\textwidth] {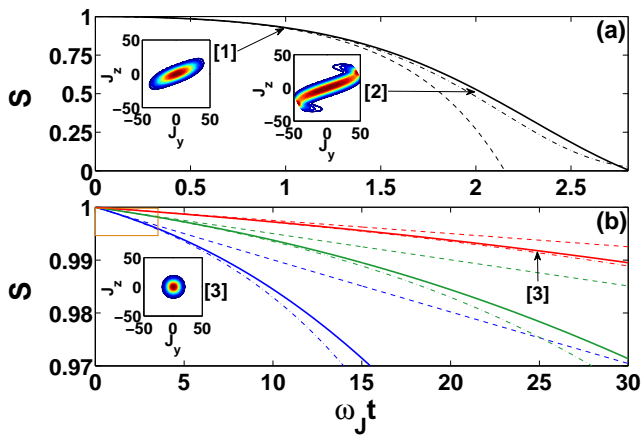}
\caption{(color online) Noise-free squeezing (a) vs.  the suppressed decay in the presence of noise (b) for the unitless time. Solid lines are numerical results, dashed lines correspond to the linearized expressions of Ref.~\cite{QZESQ1,QZESQ2}, whereas dash-dotted lines are the improved expressions of Eq.~(\ref{free}) in (a) and Eq.~(\ref{QZE}) in (b). Parameters in (a) are $u=10$, $N=100$, $D=0$, and in (b) $u=2$, $N=100$, and (from weak to strong suppression) $D=10w_J$, $20w_J$, $40w_J$. Insets show representative Husimi distributions at the marked times for the squeezed states in the absence of noise (a) and for the QZE-protected coherent state (b). The rectangular frame in the upper left corner of panel (b) denotes the time and fringe-visibility domain studied in Ref.~\cite{QZESQ1,QZESQ2}.}
\label{fig1}
\end{figure}  

{\bf One body coherence.-- }
We consider an initial coherent preparation 
that is centered at the hyperbolic point ${\bf J}_0$.
This corresponds to an $N$-particle occupation 
of the anti-symmetric superposition of the two modes \cite{Hyperbolic1,Hyperbolic2}.   
The  one-body coherence of the evolving state is 
characterized by the length of the Bloch vector ${\bf S}=\langle \hat{\bf J} \rangle/j$.      
The symmetry of the Hamiltonian (\ref{Ham}) and the initial preparation implies that $S_y=S_z=0$ so that the length of the Bloch vector is just the  {\em fringe visibility} of an experimental multiple-shot interferometric measurement.

The Wigner function of the assumed coherent state preparation 
resembles a Gaussian centered at ${\bf J}_0$
and having the angular width $r_0^2=2/N$ that corresponds 
to the minimum uncertainty of ${J_y^2+J_z^2}$. 
A squeezed state is obtained by $\eexp{\pm\Lambda}$ 
stretching along orthogonal major axes.   
The Wigner-Weyl representation of $J_z$  
is $[j(j+1)]^{1/2}\cos(\theta)$, 
with corresponding expressions for $J_x$ and $J_y$.
Accordingly, the length of the Bloch vector for a squeezed state is   
\be{0} 
S  \ &=& \ \left[1+\frac{2}{N}\right]^{1/2} |\langle\cos(r)\rangle| \ = \ \eexp{-(1/2)R^2} 
\label{fvofspread}
\\
&=& \exp\{-r_0^2 \sinh^2(\Lambda)\}~,
\label{hypspread}
\eeq
where $R^2=\langle r^2\rangle-r_0^2$
is the {\em angular spreading}.
This is equivalent to a Gaussian squeezed state approximation, where the factorization $\langle r^{2p}\rangle\approx\langle r^2\rangle^p (2p-1)!!$ is exact, allowing for the replacement of $|\langle\cos(r)\rangle|$ by $\exp(-\langle r^2\rangle/2)$.  For the dynamical squeezing under study, this approximation is valid as long as $R\ll j$. Thus, the error decreases for large $N$ and short evolution times.

\section{Loss of single-particle coherence due to squeezing}
In the absence of noise the BHH~(\ref{Ham}) induces 
pure squeezing of the initial preparation    
at the Josephson rate $w_J=\sqrt{K(NU-K)}=K\sqrt{u-1}$, 
while the angle between the squeezing axes 
is twice the value of $\Theta=\tan^{-1} (w_J/K)$ \cite{QZESQ1,QZESQ2}.
Accordingly, the total angular variance of 
the Wigner function around the hyperbolic point ${\bf J}_0$
grows initially as $\langle r^2 \rangle = \left[1+\cot^2(2\Theta)2\sinh^2(w_J t)\right]r_0^2$, 
leading to the loss of single-particle coherence as
\begin{equation}
S=\exp\left\{-r_0^2\cot^2(2\Theta)\sinh^2(w_J t)\right\}~.
\label{free}
\end{equation}
As shown in Fig.~\ref{fig1}a, Eq.~(\ref{free}) provides a good approximation for the numerically observed decay, beyond the previously used linearized form \cite{QZESQ1,QZESQ2}.

\section{Effect of noise}
Here we would like to re-consider the scenario 
that has been analyzed in \cite{QZESQ1,QZESQ2}. 
Using the analogy to the standard QZE, 
the result that has been obtained there 
was an exponential decay 
\beq
S=\exp[-r_0^24D_w t]~, \ \ \ D_w=[\cot^2(2\Theta)]\frac{w_J^2}{8D}~.
\label{QZEL}
\eeq
As seen in Fig.~\ref{fig1}b, while this expression is accurate for the first few Josephson periods, it fails  on longer time-scales. Long time accuracy may be improved using the semi-classical strategy 
of the previous paragraph.  Thus, instead of applying the QZE sequence of projections directly to $S$ as done in \cite{QZESQ1,QZESQ2}, we apply it here to the angular variance $\langle r^2\rangle$.
One may visualize a sequence of squeezing intervals 
of duration $t_D=1/(2D)$ wherein $\langle r^2\rangle$ 
grows as $\langle r^2\rangle_{t+t_D}=[1~+~2\cot^2(2\Theta)\sinh^2(w_J t_D)]\langle r^2\rangle_t$ 
before being reset by the noise. In the limit where $t_D\ll t_J\equiv 1/w_J$,  
the spreading within each interval is quadratic in time, 
so that $\langle r^2\rangle_t=r_0^2\exp[\cot^2(2\Theta)(w_J^2/D)t]$.  
Consequently 
\begin{equation}
S=\exp\left\{-\frac{r_0^2}{2}\left[\exp\left(8D_wt\right)-1\right]\right\}~.
\label{QZE}
\end{equation}
Comparison of Eq.~(\ref{QZEL}) and Eq.~(\ref{QZE})  to the full numerical evolution (Fig.~\ref{fig1}b)  demonstrates great improvement over the short time result of  Ref.~\cite{QZESQ1,QZESQ2}. The simple exponential of Eq.~(\ref{QZEL}) is only valid for $t\ll t_{QZ}\equiv 1/D_w$ 
where it can be approximated by a linear function. 
In comparison, Eq.~(\ref{QZE}) is valid for $t<t_{QZ}\log(N)$, 
after which $S$ decreases significantly below unity
and the Gaussian approximation no longer holds.

\begin{figure}
\centering
\includegraphics[width=0.48\textwidth] {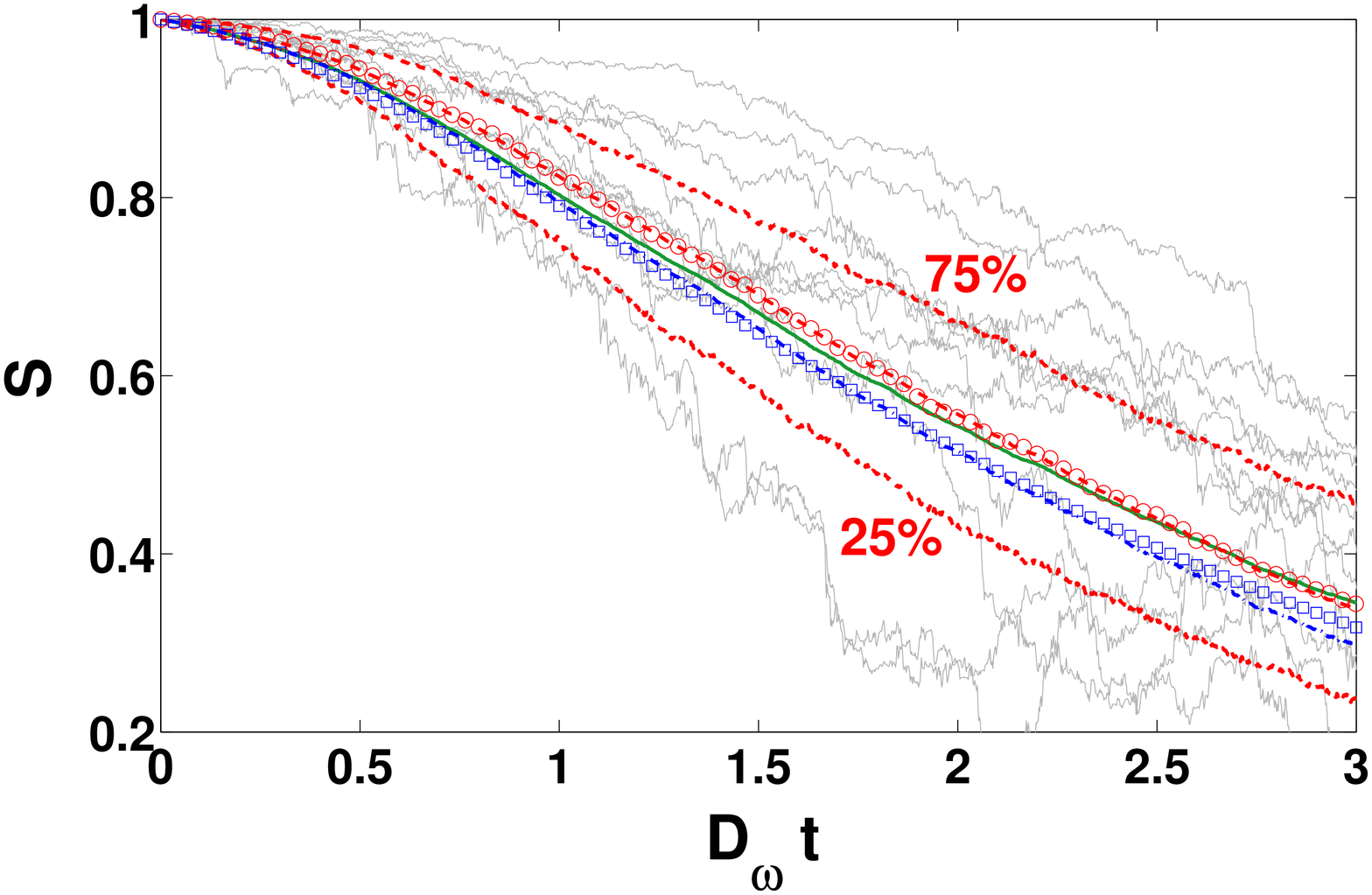}
\caption{(color online) Fringe visibility dynamics for erratic driving for $N=50$, $u=2$, $D/w_J=37.5$, and unitless time. Single realizations of erratic driving are marked by gray lines. The mean over $2000$ realizations (solid green) lies in between the median value (dashed red) and the true average over an infinite number of realizations, which is equal to the propagation of the master-equation (\ref{master})  (dash-dotted blue). Symbols correspond to the extraction of the median (circles) and average (squares) from the Gaussian approximation of the squeezing parameter $\Lambda$ distribution, according to Eq.~(\ref{median}) and Eq.~(\ref{mean}), respectively. Dotted red lines mark the 25th and 75th percentiles.}
\label{fig2}
\end{figure}  

\section{Erratic Driving vs. Noise} 
While Eq.~(\ref{free}) and Eq.~(\ref{QZE}) dramatically improve our quantitative understanding of the many-body QZE, far beyond the previously studied regime, our main focus here is to challenge the paradigm  of replacing {\em quantum noise} by deterministic {\em erratic driving}.  Erratic driving means that the Hamiltonian is time dependent due to some deterministic but fluctuating~$f(t)$. An experimentalist can repeat one experiment many times with exactly the same~$f(t)$, and determine the final quantum state. The experimentalist can also repeat the experiment with different realizations of~$f(t)$ and accumulate {\em statistics}. 

By contrast, a noisy process as described by Eq.~(\ref{master}) can be viewed as arising from~$f(t)$ realizations that are induced by a bath. These realizations are not under experimental control: the individual $~f(t)$ 
cannot be reproduced from run to run. The best measurement the experimentalist can do already yields 
a density matrix $\rho$, which we may call the {\em average}. In effect, it is Nature, 
rather than the experimentalist, who averages over $f(t)$. While the experimentalist considers  
individual realizations of~$f(t)$, Nature averages over {\em all} realizations. 

It is thus clear that in the case of erratic driving we should consider the 
{\em statistics} of $S=S[f]$, while in the case of a noisy driving only 
the averaged $S$ (i.e., the $S$ of the mixed state) has a physical meaning, 
as implied by the master equation.     

Alternatively, erratic driving is aiming to emulate quantum noise by realizations that sample the ensemble of all possible paths. The naive expectation would be that for a reasonably large number of such realizations one would obtain a typical value which coincides with the true average. However, we show below that due to the interplay of Gaussian randomization and hyperbolic amplification, rare events which are missed by erratic driving play an important role in determining the final (typical) outcome of the squeezing in the presence of noise. Consequently, erratic driving sampling will typically differ substantially from the ideal average, even when the number of realizations is large.

\begin{figure}
\centering
\includegraphics[width=0.5\textwidth] {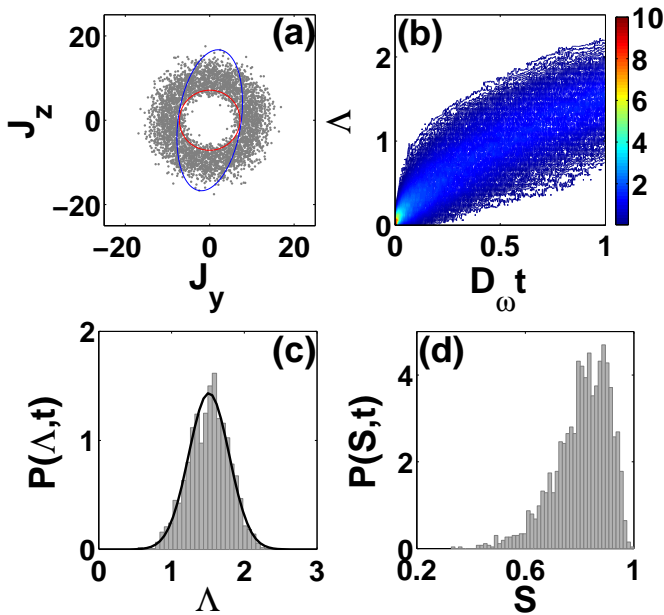}
\caption{(color online) Statistics of erratic driving: (a) Distribution of realizations. Each squeezed state is represented by a pair of points along its long principle axis, separated by the principle-axis variance $\Delta_+$, as illustrated for two representative realizations; (b) evolution of the squeezing parameter $\Lambda$ distribution; (c) Gaussian distribution of $\Lambda$; (d) log-wide distribution of the fringe visibility. Parameters are the same as in Fig.~\ref{fig2}.  Panel (a) is taken at $D_w t=0.35$ whereas panels (c) and (d) are taken at $D_w t=0.5$.}
\label{fig3}
\end{figure}  

\section{Effect of erratic driving}

Fig.~\ref{fig2} displays the time dependence of~$S$ for a few representative 
realizations of erratic driving, out of a large sample of 2000 random scenarios. 
The mean, median, 25th and 75th percentiles, and the true average (from the master equation simulation) are indicated as well.  As shown, the sample mean taken over the entire ensemble 
deviates from the ``true" master equation result, and lies between the median and the true average. 

In order to explain the observed difference between erratic driving and noise, we perform a statistical analysis of the evolving ensemble of squeezed states under erratic driving. In Fig.~\ref{fig3}a  we show the distribution of  the squeezing axis direction and the degree of squeezing, with two extreme states. Each Gaussian squeezed state is represented by a pair of points along its long principle axis. The distance between the points is $\Delta_+$, which is two times the square root of the long axis variance. The latter obtained by the diagonalization of the variance matrix $\Delta_{ij}=\langle \Ji\Jj+ \Jj\Ji\rangle/2-\langle\Ji\rangle\langle\Jj\rangle$ with $i,j=y,z$. The empty internal ring corresponds to the minimal coherent state variance. We note that the principle axis direction is completely randomized with rare events of repeated stretching (the distant points).

The resulting $S$ distribution is log-wide as shown in Fig.~\ref{fig3}d, with its median significantly smaller than its average. This log-normal statistics is explained as follows: for a given realization of $f(t)$ 
the wavepacket undergoes a sequence of squeezing operations. Dividing the time into intervals 
of size $t_D$, one realizes that the squeezing operations are uncorrelated, and can be regarded as a 
random sequence of stretching and un-stretching steps. Accordingly, the accumulated squeezing parameter $\Lambda$ 
is a sum of uncorrelated variables, and according to the central limit theorem it should have
a normal distribution. From Eq.~(\ref{hypspread}) it follows that $S$ will have log-wide distribution.     

\begin{figure}
\centering
\includegraphics[width=0.48\textwidth] {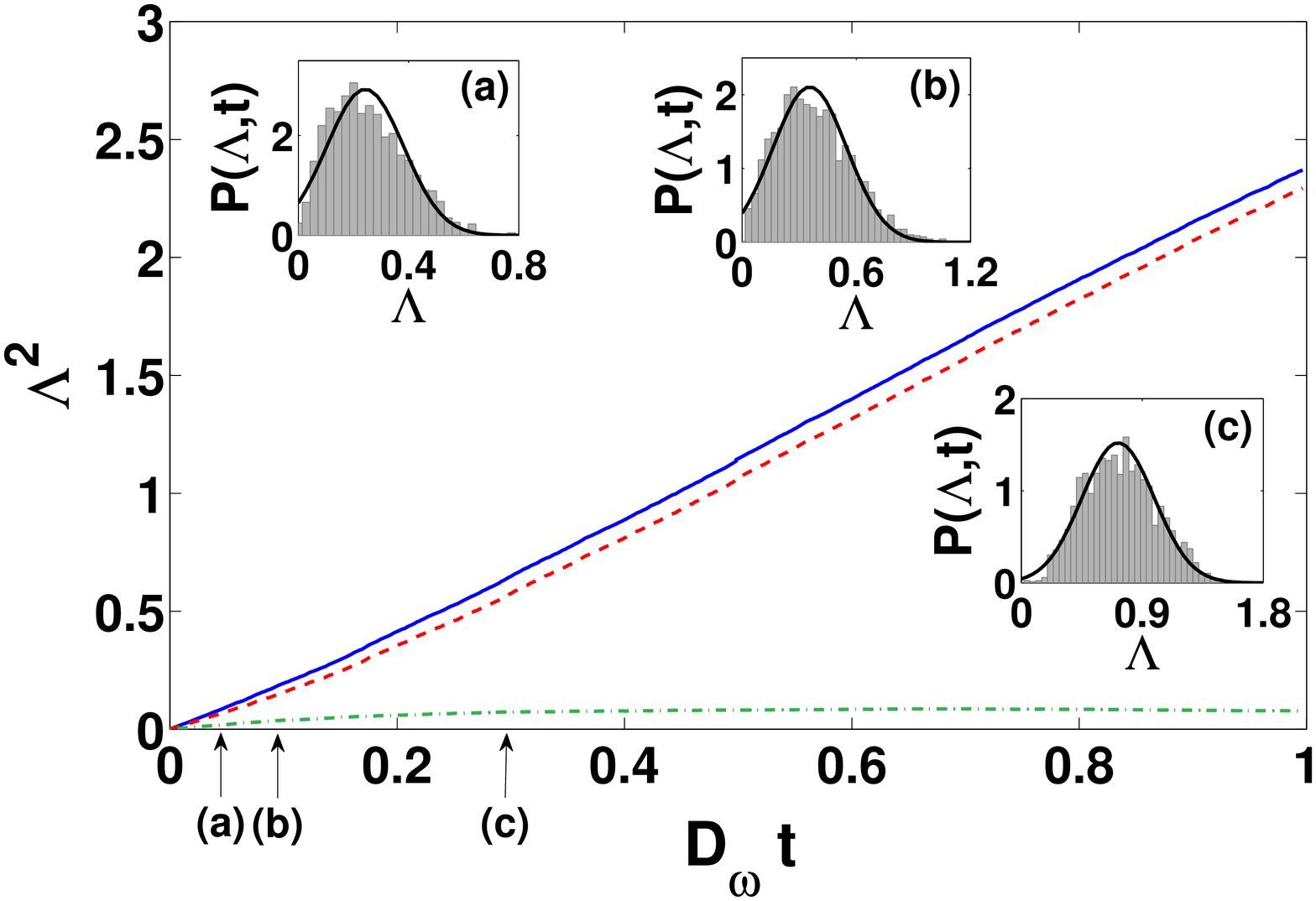}
\caption{(color online) Evolution of the mean squared $\mu^2$ (dashed red) and the variance $\sigma^2$ (dash-dotted, green) of the squeezing parameter distribution. The solid blue line denotes $\langle \Lambda^2\rangle=\mu^2+\sigma^2$. Insets depict the squeezing parameter distribution for 2000 realizations at the marked times, with superimposed Gaussian fits (black lines). Parameters are the same as in Fig.~\ref{fig2} and Fig.~\ref{fig3}}
\label{fig4}
\end{figure}

We can deduce the effective squeezing parameter $\Lambda$ for each realization from its single-particle coherence $S$ by inverting Eq.~(\ref{hypspread}). The time evolution of the deduced $\Lambda$ distribution 
is shown in Fig.~\ref{fig3}b, and a representative cross section  is plotted in Fig.~\ref{fig3}c. 
As expected, after a short transient of unfolding the squeezing parameter distribution takes a Gaussian form, 
for which we find the mean $\mu=\langle \Lambda\rangle$ and variance $\sigma^2=\langle \Lambda^2\rangle-\mu^2$. 
The obtained $\mu^2(t)$ and $\sigma^2(t)$ are plotted as a function of time in Fig.~\ref{fig4}, 
along with three representative insets for the $\Lambda$ distribution.

Having characterized the $\Lambda$ distribution by its $\mu$ and $\sigma$, 
we can now go back and deduce the expected values for the median 
and the average of~$S$. The median value is obtained by substitution of the 
prevalent squeezing parameter $\mu$ into Eq.~(\ref{hypspread}),
\begin{equation}
S_{med}=\exp\left\{-r_0^2\sinh^2(\mu)\right\}~,
\label{median}
\end{equation}
whereas the mean value is found by averaging,
\begin{eqnarray}
\label{mean}
S_{avg}&\approx& \exp[-r_0^2\langle\sinh^2(\Lambda)\rangle]\\
~&=&\exp\left\{-\frac{r_0^2}{2}\left[{\rm e}^{2\sigma^2}\cosh(2\mu)-1\right]\right\}~.\nonumber
\end{eqnarray}
As shown in Fig.~\ref{fig2}, substitution of $\mu$ and $\sigma$ from Fig.~\ref{fig4} into Eq.~(\ref{median}) and Eq.~(\ref{mean}) gives excellent agreement with the median and {\em true} average of the $S$ distribution. 

To conclude, small sampling errors of the normal $\Lambda$ distribution
correspond to miss-sampling of the tail of the log-wide distribution 
of the spreading~$R$, whose median is distinct from its average. 
Since the average $S$ is strongly affected by the tails, we end up producing large errors. 
However, the miss-sampled tails can be properly deduced from 
a Gaussian approximation for the squeezing-parameter distribution, 
so as to overcome the sampling issue and get a prediction 
for the true average. We note that without performing this procedure,  
the average obtained by an experimentalist over many realizations 
of erratic driving is likely to reflect the {\em median}, which is the typical value, 
rather than the true average of the distribution.

%
\section{Summary}
We have studied the process of squeezing around the hyperbolic fixed point of the two site Bose-Hubbard model, in the presence of intense noise. We have greatly extended the quantitative understanding  of the observed Quantum Zeno effect \cite{QZESQ1,QZESQ2} and investigated one of the principle paradigms in the theory of quantum noise, namely the replacement of an ideal quantum bath by deterministic erratic driving. We have shown that the interplay of diffusive quantum noise and hyperbolic squeezing results in log-wide statistical distributions of variances and fringe visibilities, so that their mean is different than their typical value. Consequently, we find that the fringe-visibility dynamics in a typical erratic driving scenario will differ from that obtained by coupling to an ideal quantum bath.


{\bf Acknowledgments.-- }
We thank James Anglin for fruitful discussions. This research was supported by the Israel Science Foundation (grant Nos. 346/11 and 29/11) and by grant No. 2008141 from the United States-Israel Binational Science Foundation (BSF).



\begin{thebibliography}{99}
\bibitem{rA}
In the numerical calculation, we discretize the time, dividing it into short steps  
of duration~$\tau$. The noise is realized by introducing a random 
rotation $\exp(\pm i\Delta\Phi \Jx)$ at the end of each step.     
This random walk process corresponds to erratic driving with the diffusion 
coefficient $D=(\Delta\Phi)^2/(2\tau)$.
\end{thebibliography}
\end{document}